# Morphological Computing as Logic Underlying Cognition in Human, Animal, and Intelligent Machine

Gordana Dodig-Crnkovic [1,2,*]

[1] Department of Computer Science and Engineering, Chalmers University of Technology, 412 96 Gothenburg, Sweden
[2] School of Innovation, Design and Engineering, Computer Science Laboratory, Mälardalen University, Sweden

## Abstract

This work examines the interconnections between logic, epistemology, and sciences within the Naturalist tradition. It presents a scheme that connects logic, mathematics, physics, chemistry, biology, and cognition, emphasizing scale-invariant, self-organizing dynamics across organizational tiers of nature. The inherent logic of agency exists in natural processes at various levels, under information exchanges. It applies to humans, animals, and artifactual agents. The common human-centric, natural language-based logic is an example of complex logic evolved by living organisms that already appears in the simplest form at the level of basal cognition of unicellular organisms. Thus, cognitive logic stems from the evolution of physical, chemical, and biological logic.

In a computing nature framework with a self-organizing agency, innovative computational frameworks grounded in morphological/physical/natural computation can be used to explain the genesis of human-centered logic through the steps of naturalized logical processes at lower levels of organization. The Extended Evolutionary Synthesis of living agents is essential for understanding the emergence of human-level logic and the relationship between logic and information processing/computational epistemology.

We conclude that more research is needed to elucidate the details of the mechanisms linking natural phenomena with the logic of agency in nature.

## Introduction

Morphological computing and cognition refer to the study of how physical structures and processes in biological systems contribute to information processing and cognitive functions. The idea is that an organism or a robotic system's physical shape and material properties define its cognitive abilities and information processing capabilities. This concept is inspired by how biological systems, such as humans and animals, use their body morphology to process information and solve problems more efficiently than purely symbolic-level computational approaches.

Connecting morphological computing and cognition to logic involves understanding how the physical structure and properties of an organism or a robotic system can be used to represent and manipulate logical information. Practical logic and logic in action are related concepts that emphasize the real-world applicability of logic, focusing on how logical reasoning and operations can be used to solve problems and make decisions in everyday life.

By connecting morphological computing and cognition to practical logic and logic in action, we aim to develop new ways of designing and building artificial intelligence systems, robots, and cognitive models that are more closely aligned with the principles found in natural biological systems.

This interdisciplinary approach draws on insights from fields such as biology, computer science, neuroscience, and philosophy.

Some possible applications of this connection include:

- developing AI systems that use morphological computing principles to make more efficient use of computational resources and better resemble human cognitive abilities.
- creating new cognitive models and theories that integrate morphological computing and logic to better understand human and animal cognition.
- investigating how the morphology of biological systems can inform the design of new materials and devices for information processing and problem-solving.
- designing robots with physical structures that can efficiently process information and solve problems by exploiting their body morphology.
- developing logical theory in the direction of "logic in reality" (Brenner 2008)(Brenner 2012)

Generally, connecting morphological computing and cognition to logic, practical logic, and logic in action is an interdisciplinary research area with the potential to advance our understanding of cognition and inform the development of more efficient and biologically inspired artificial intelligence systems and robots.

## Practical logic of Gabbay and Woods

In "Agenda Relevance: A Study in Formal Pragmatics," Dov M. Gabbay and John Woods present their work "A Practical Logic of Cognitive Systems" (Gabbay and Woods 2003), where they introduce the concept of "practical logic." They explain that this type of logic is rooted in pragmatics, which has historically been a branch of the theory of signs involving *non-trivial and irreducible references to agents – entities that receive and interpret messages.*

The authors define a cognitive agent as a being capable of perception, memory, belief, desire, reflection, deliberation, decision, and inference. A practical cognitive system is a cognitive agent that is an individual. The practical logic they describe offers a specific kind of description of such a practical cognitive system.

They clarify how a pragmatic theory of reasoning includes non-trivial and irreducible references to cognitive agents. If we consider cognitive agents as specific types of information processors, then a pragmatic theory of cognitive agency will offer descriptions of these information processors.

Given that "logic is a principled account of certain aspects of practical reasoning", it is also intrinsically pragmatic. The proper province of logic in practical reasoning, according to Gabbay and Woods, lies in operational arrangements. (Gabbay and Woods 2003)(Benthem et al. 2001)

Johan van Benthem, in his book "Logical Dynamics of Information and Interaction" (Benthem 2011), highlights that practical logic is part of a pragmatic theory addressing the necessary aspects of practical cognitive agency *at both linguistic and sublinguistic levels.* For this, "a suitably flexible notion of information is necessary". Benthem emphasizes that his approach to practical logic, shares similarities with the "dynamic turn" in logic research. This approach focuses on the intersection of cognitive science and experimental studies to *understand the underlying psychological and neurological realities of human information processing and cognition* (Benthem et al. 2001), p.5.

Logic, Rationality, Interaction, and Naturalization of Logic

The "International Workshop on Logic, Rationality and Interaction" (Fenrong 2023) (LORI 2021) promotes the idea *that studying information primarily involves studying information exchange, acknowledging the inherently relational nature of information.* The 2007 LORI workshop explored new interfaces like epistemic studies of rational behavior in games, social software, and the role of interaction in natural language. The field of pragmatics, which focuses on the actual use of language between agents, has become the primary research focus in this context. Game theory, particularly evolutionary games, is being employed to address various pragmatic issues, including the development of linguistic conventions. Information exchange is a form of interaction where agents act together in strategic ways.

These new perspectives on logic, rationality, and interactions between agents have led to contacts between logic and game theory, bringing a new set of disciplines into the scope of logic such as economics, and the social sciences. Alexandru Baltag and Sonja Smets explore this idea further in their book "Johan van Benthem on Logic and Information Dynamics", (Baltag and Smets 2014).

The evolution of this field has led to a shift from a human-centric perspective in 2007 to a cognitive and intelligent agent-centric perspective in 2022, considering both living agents and artificial agents. This progression suggests a generalization of the inherently relational nature of information and logic across various levels of cognitive/intelligent systems.

Among the early proponents of naturalization in logic, John Dewey has a notable place with his "Essays in experimental logic" which provided an account of logic based on the natural sciences (Dewey 1916). Dewey indirectly suggested that the relationship between the world and logic should also extend from science to logic. In this view, logic appears to play a secondary role in relation to science, particularly physics. This perspective aligns with Quine's position on epistemology, (Quine 1969), ontology, and other philosophical branches, which he believed should also be subordinate to science. Quine specifically argued against the existence of an a priori epistemology or philosophy in general, implying that there is no fully a priori logic either.

Connecting the dots between Logic and Epistemology, one might see this paper as following the steps of Quine's "Epistemology Naturalized" (Quine 1969) in which Quine claims:
"If all we hope for is a reconstruction that links science to experience in explicit ways short of translation, then it would seem more sensible to settle for psychology. Better to discover how science is in fact developed and learned than to fabricate a fictitious structure to a similar effect." (Quine 1969) (p. 78). Epistemology, according to Quine, is a part of psychology and thus of natural science.

However, the ambition of the present work does not end on the level of psychology, which translates to cognitive science in this context. We search for the genesis of logic even deeper, relating all of knowledge production to the mechanisms of natural processes extrinsic to the cognizing agent, as well as mechanisms intrinsic to the agent, in the continuous interaction.

Another, a more recent prominent proponent of the naturalization of logic, Jan Woleński in the articles, "Logic in the Light of Cognitive Science", and "Naturalism and Genesis of Logic" builds on the experiential character of knowledge that can be understood phylogenetically which allows us to investigate the genesis of logic through the lenses of evolutionary theory, (Woleński 2012)(Woleński 2012).

Similarly, John Woods in his *Errors of Reasoning: Naturalizing the Logic of Inference* (Woods 2013) argues for the necessity of naturalizing logic.

Lorenzo Magnani made 2018 a proposal arguing for the urgent need for naturalizing logic (Magnani 2018). Among the recent naturalization of the logic research Paul Thagard studied how knowledge of mechanisms enhances inductive inference and shows how knowledge about real mechanisms contributes to "generalization, inference to the best explanation, causal inference, and reasoning with probabilities.", (Thagard 2021). All of the mentioned proposals concern human-level logic.

The work of Joseph Brenner is also along the lines of natural logic, or in his own words, Logic in Reality, (Brenner 2008)(Brenner 2012). Related work on naturalization by Terrence Deacon can also be seen as a proposal on the logic of (informational) nature (Deacon 2011), (Brenner 2012).

Within the naturalistic framework, Thomas Parr, Giovanni Pezzulo, and Karl J. Friston, recently presented generative models for sequential dynamics in the active inference (Parr, Pezzulo, and Friston 2022) *where the dynamics of physical systems minimize a surprise, or equivalently, its variational upper bound, free energy. The free energy principle is based on the Bayesian idea of the brain as an "inference engine." Under the free energy principle, systems pursue paths of least surprise, or, minimize the difference between predictions based on their model of the world and associated perception. In short, this approach covers the steps of inference generation and testing in the human brain.*

Gianfranco Basti presents the perspective of a physicist, in the article "The Philosophy of Nature of the Natural Realism. The Operator Algebra from Physics to Logic", (Basti 2022). Andrée Ehresmann, and Jean-Paul Vanbremeersch, also in the framework of Contemporary Natural Philosophy, argue for an integral study of living systems: evolutionary multi-level, multi-agent, and multi-temporary self-organized systems, such as biological, social, or cognitive systems. Their Memory Evolutive Systems (MES) approach is based on a 'dynamic' Category Theory, proposing an info-computational model for living systems, (Ehresmann and Vanbremeersch 2019).

From the epistemological point of view, the foundations of the new Naturalism are outlined by (Ladyman, Ross, and Spurrett 2007) in the book "Every Thing Must Go: Metaphysics Naturalized", while (Dodig-Crnkovic 2007) proposed the info-computationalist approach as a method to achieve this naturalization.

The strictly science-based approaches to naturalization can be contrasted with phenomenological frameworks, which traditionally question scientific framing. Despite this, Maurita Harney presents an argument for naturalizing phenomenology as a philosophical necessity (Harney 2015). She explains the historical roots of phenomenology's skeptical attitude towards science, evident in "Husserl's dismissal of 'the scientific attitude', Merleau-Ponty's differentiation from the scientifically objectified self, and Heidegger's critique of modern science." However, recent advancements in neuroscience have created new possibilities for collaboration between phenomenology and cognitive science, prompting a reevaluation of science and its underlying assumptions. Central to this is the reimagining of nature as encompassing meanings and the mind. Compare to Valentino Braitenberg's notion "Information - der Geist in der Natur" (Braitenberg 2011)

In the context of phenomenological naturalism, meaning should be understood through a relational ontology rather than an atomistic one, and characterized by dynamic, process-oriented properties instead of static, substance-based qualities.

## Cognition in Nature and Artifacts as Computation of Information

Going even further in the naturalization process, ascribing logic not only to human (and AI based on human logic) agents but generalizing it further to natural processes in the sense of a self-generating loop of recursive logic hierarchies, our goal is to connect human-centric, human language-based logic (grounded in cognition) with natural logic of not only physical but also chemical, biological and cognitive processes in all living agents.

An info-computational framework for the analysis of cognition and intelligence, natural and artificial, is a foundation for the study of information processing/ computational phenomena.

In the study of cognition in nature, a two-way learning process takes place: from a theoretical and experimental study of natural systems to a constructive study of artifacts (such as deep neural networks, machine learning, and robots) and from increasingly sophisticated artifacts back to models and theories of structures and behaviors of natural systems (such as brains, swarms, or unicellular organisms), (Rozenberg and Kari 2008).

At the time when the first models of cognitive architectures (inspired by human cognition) were proposed, some forty years ago, the understanding of cognition, embodiment, and evolution was substantially different from today.

## Logics in the Wild

In the first chapter of their book on Biochemistry, "The Facts of Life", Reginald Garrett and Charles Grisham argue: "Chemistry is the logic of Biological Phenomena", based on the insight that biomolecules are informational carriers that provide structure and generative rules for biological cells (Garrett and Grisham 2023). In a more detailed computational approach, (Foulon et al. 2019) present a language for molecular computation.

If chemistry is the logic of biology, what is the logic of chemistry? In search for the role of logic in knowledge production, decision-making, and agency in nature, we propose the following scheme, in a recursive (circular) manner, starting with logic:

- Logics is the intrinsic foundation of mathematics.
- Mathematics is the intrinsic logic of physics.
- Physics is the intrinsic logic of chemistry.
- Chemistry is the intrinsic logic of biology.
- Biology is the intrinsic logic of cognition.
- Cognition generates (agent's intrinsic) logic in the first place.
  An agent's intrinsic logic is a basis for creating extrinsic/externalized/shared logic. Go to 1.

On a systemic view, nature can be seen through scale invariant self-organizational dynamics of energy/matter at the hierarchy of levels of organization, (Kurakin 2011).

The logic of living beings refers to the principles and processes that govern the behavior and functioning of organisms. It encompasses the various mechanisms through which living beings interact with their environment, survive, reproduce, and evolve over time, which unfolds through a hierarchy of levels as described above.

## Info-Computational Nature

The Computing Nature (Naturalist Computationalism, Info-Computationalism) framework makes it possible to describe all cognizing agents (living organisms and artificial cognitive systems) as informational structures with computational dynamics, (Dodig-Crnkovic 2022b)(Dodig-Crnkovic 2022a)(Dodig-Crnkovic and Giovagnoli 2013).

Even Alan Turing's pioneering work on computability and his ideas on morphological computing can be seen as a position in Natural Philosophy, (Hodges 1997). Turing's natural philosophy differs importantly from Galileo's belief that the book of nature is written in the language of mathematics (Galilei 1623). Computing is more than a language (description) of nature as computation also produces real-time physical behaviors, a driving mechanism of natural processes.

This article puts the computational approaches into a broader context of natural computation, where information dynamics is not only found in human communication and computational machinery but also in the entire nature. Information is understood as representing the world (reality as an informational web) for a cognizing agent, while information dynamics (information processing, computation) realizes physical laws through which all the changes of informational structures unfold.

Computation as it appears in the natural world is more general than the human process of calculation modeled by the Turing machine. Natural computing is epitomized through the interactions of concurrent, in general, asynchronous computational processes which are adequately represented by what Abramsky names "the second generation models of computation" (Abramsky 2008) which we argue to be the most general representation of information dynamics. Conceptualizing the physical world as an intricate tapestry of information networks evolving through processes of natural computation helps to make more coherent models of nature, connecting non-living and living worlds. It presents a suitable basis for incorporating current developments in understanding biological/ cognitive/ social systems as generated by the complexification of physicochemical processes through the self-organization of molecules into dynamic adaptive complex systems by morphogenesis, adaptation, and learning - all of which are understood as information processing. (Dodig-Crnkovic 2014)

## Learning from Nature Within Info-Computational Framework Requires Updates of Definitions

As explained in (Dodig-Crnkovic 2022b), in order to construct a naturalist explanation of the logical basis of cognition and the cognitive basis of logic, some definitions have been generalized, that we should keep in mind.

**Information** = structure for a cognizing agent. It means not only news and artifacts in our human civilization that are used to transmit data and knowledge, but also similar structures utilized by other living organisms (cognitive agents), even the simplest ones like bacteria.

**Computation** = dynamics of information. It is taken to be any process of information transformation, that leads to behavior, and not only those processes that we currently use to calculate, manually or with machinery.

**Cognition** = life (for a living organism). It is the ability to learn from the environment and adapt so as to survive as individuals and species, for which organisms use the information and its processing - computation. Intelligence as the capacity for problem-solving can be found in all organisms as they all possess cognition. The cognition of the simplest organism, a single cell, is called basal cognition. (Lyon et al. 2021a)(Lyon et al. 2021b)(Levin et al. 2021a) Understanding cognition and intelligence based on biological mechanisms is only possible if we see it in the context of evolution. Finally, cognition and intelligence in artifacts is the project of building machines mimicking human capabilities.

**Evolution** is understood as *an extended evolutionary synthesis* formulation of evolutionary theory, which is the interpretation of the theory of evolution based on the newest scientific knowledge about life and its changes, emphasizing fundamental mechanisms of constructive development and reciprocal causation with the environment. (Laland et al. 2015) (Ginsburg and Jablonka 2019)

## Actor Model of Concurrent Distributed Computation

"In the Actor Model [Hewitt, Bishop, and Steiger 1973; Hewitt 2010], computation is conceived as distributed in space, where computational devices communicate asynchronously, and the entire computation is not in any well-defined state. (An Actor can have information about other Actors that it has received in a message about what it was like when the message was sent.) Turing's Model is a special case of the Actor Model.", (Hewitt 2012).

Hewitt's "computational devices" are conceived as computational agents – informational structures capable of acting on their own behalf.

## Computing Cells: Self-generating Systems

Complex biological systems must be modeled as self-referential, self-organizing "component-systems" (Kampis 1996) which are self-generating and whose behavior, though computational in a general sense, goes far beyond Turing machine model repertoire of behaviors.

"a component system is a computer which, when executing its operations (software) builds a new hardware... [W]e have a computer that re-wires itself in a hardware-software interplay: the hardware defines the software, and the software defines new hardware. Then the circle starts again." (Kampis 1991) p. 223

Computing understood as information processing is closely related to natural sciences; it helps us recognize connections and levels of organization in natural sciences and provides a unified approach for modeling and simulating both living and non-living systems. (Dodig-Crnkovic 2011) (Dodig-Crnkovic 2017)

One of the important aspects of natural computing is its computational efficiency. The Turing Machine model of computation is not resource-aware, unlike living systems that are constantly optimizing their resource use. Ihor Lubashevsky in the book "Physics of the Human Mind" explicitly addresses those aspects (Lubashevsky 2017).

Our present-day von Neumann computing architecture has bottlenecks, as information processor and memory are separate. Memristors as a biomimetic solution combine memory and processor and avoid von Neumann bottlenecks. The proposed solutions to the problem of our energy-consuming contemporary computing machinery are typically inspired by nature. E.g. Kevin M. Passino proposes biomimicry for optimization, control, and automation, (Passino 2005). Melanie Mitchell addresses the question of unconventional, biological computation in the (Mitchell 2012). More on natural computing can be found in (Rozenberg and Kari 2008)(Rozenberg, Bäck, and Kok 2012).

## Cognition = Life: Agency-based Hierarchies of Levels. The World as Information for an Agent

Cognition in living systems/agents constitutes life-organizing, life-sustaining goal-directed processes, (Stewart 1996) (Maturana and Varela 1980) (Maturana and Varela 1992)(Marijuán, Navarro, and del Moral 2010), while in artifactual systems, cognition is an engineered process based on sensors, actuators, and computing units designed to mimic biological cognition (bio-mimetic design). Cognitive architectures generated by natural (morphological) computation are realized in a specific substrate of matter/energy self-organized in living cells (Dodig-Crnkovic 2012).

## Basal Bio-cognition

The work of Michael Levin suggests a broad range of applications for nature-inspired cognitive architectures based on biological cognition connecting genetic networks, cytoskeleton, neural networks, tissues/organs, and organisms with the group (social) levels of information processing. Levin shows how biology has been computing through *somatic memory* (information storage) and *biocomputation/decision-making* in *pre-neural bioelectric networks*, before the development of neurons and brains. (Levin et al. 2021a)

Insights from bio-cognition can help the development of new AI platforms, applications in targeted drug delivery, regenerative medicine and cancer therapy, nanotechnology, synthetic biology, artificial life, and much more.

Recent research finds that "*cognitive operations we usually ascribe to brains—sensing, information processing, memory, valence, decision making, learning, anticipation, problem-solving, generalization and goal-directedness—are all observed in living forms that don't have brains or even neurons.*" (Levin et al. 2021)(Levin et al. 2021b).

Thus, we generalize cognition a step further, to include all living forms, not only those with nervous systems. It can be useful for artificial systems that need a level of intelligence but not the human level, such as nano-bots or different elements of IoT.

## Bacterial Cognition and Bacterial Chemical Language

For example, symbolic information processing can be found both on the level of human languages, but also on the level of chemical languages used by bacteria, as Bassler (Bacterial quorum sensing) (Ng and Bassler 2009) and Ben-Jacob (Bacterial Know How: From Physics to Cybernetics) (Ben-Jacob 2008)(Ben-Jacob, Becker, and Shapira 2004)(Ben-Jacob 2009) have described.

A framework of natural cognition based on info-computation in living agents enables the unification of natural and artificial cognition and intelligence. Cognition in nature is a manifestation of biological processes in all living beings, that subsume chemical and physical levels.   Intelligence is considered a problem-solving ability at different levels of organization.

## Learning from Basal Cognition on the Continuum of Natural Cognition

The concept of biological computation implies that living organisms perform computations, and that as such, abstract ideas of information and computation may be key to understanding biology.

> *Apart from the human brain with a nervous system, cognition can be found in somatic cells, non-human organisms with a nervous system, non–neuronal subsystems in humans such as the immune system.* (Dodig-Crnkovic 2021)

## Systems Biology View

Individual cells are capable of making decisions based on their internal condition and environment, but the exact mechanism for reliable decision-making remains unknown. Investigations of (Kramer, Castillo, and Pelkmans 2022) focus on the information-processing abilities of human cells by measuring various signaling responses and cellular state indicators. The signaling nodes within a network exhibited adaptive information processing, resulting in diverse growth factor reactions and allowing nodes to gather partially unique information about the cell's condition. In combination, this provides individual cells with a substantial information processing capacity to accurately determine growth factor concentrations in relation to their cellular state and make decisions accordingly. According to (Kramer et al. 2022) the coevolution of heterogeneity and complexity within signaling networks may have contributed to the development of precise and context-sensitive cellular decision-making in multicellular organisms.

## Evolution Provides Generative Mechanism for Increasingly Complex Cognition and Logic

New insights about cognition and its evolution and development in nature from cellular to human cognition can be modeled as natural information processing/ natural computation/ morphological computation. In the info-computational approach, evolution in the sense of extended evolutionary synthesis is a result of interactions between natural agents, cells, and their groups.
Evolution provides a generative mechanism for the emergence of increasingly more competent living organisms with increasingly complex natural cognition and intelligence which are used as a template for their artificial/computational counterparts.

## Natural Computation Driving Evolution

The view of computing nature abounds in recent literature (Zenil 2012) (Dodig-Crnkovic 2022a)(Dodig-Crnkovic and Giovagnoli 2013) (Chaitin 2018). Its basis is an unconventional generalized view of computation, natural computation. As Greg Chaitin explains:

> *"And how about the entire universe, can it be considered to be a computer?  Yes, it certainly can, it is constantly computing its future state from its  current state, it's constantly computing its own* *time evolution!* And as I believe Tom Toffoli pointed out, actual computers like your PC just hitch a ride on this universal computation!" (Chaitin 2007)

A similar approach can be found in (Dodig-Crnkovic 2007a) as well as elaborated in the article "What is Computation? (How) Does Nature Compute?" by David Deutsch in (Zenil 2012)

## Evolution as Learning. Cognition as a Driver of Evolution and Evolution as a Driver of Cognition in Living Organisms. From Modern Evolutionary Synthesis to Extended Evolutionary Synthesis

A contemporary emerging, broader perspective on cognition not only recognizes humans as cognitive agents but includes all living organisms (Dodig-Crnkovic 2022a). It enables a fresh insight into the mechanisms of the evolutionary process. Evolution is now seen as a cognition-driven process, influenced by targeted genome editing and natural intrinsic cellular engineering (Miller, Enguita, and Leitão 2021). Torday and Miller explore this cognition-centric view of evolution particularly within epigenetic evolutionary biology (Torday and Miller 2020). Along the same lines, Miller, Enguita, and Leitão explain that: "Cognition-Based Evolution argues that all of biological and evolutionary development represents the perpetual auto-poietic defense of self-referential basal cellular states of homeostatic preference. The means by which these states are attained and maintained is through self-referential measurement of information and its communication." (Miller et al. 2021) Also, Baluška, Miller, and Reber studied the cellular and evolutionary perspectives on cognition, tracing the path from unicellular to multicellular organisms. Their research focuses on the evolutionary origins of cells, starting from unicellular organisms in their progression towards multicellularity. In this view, his evolution centers on two core self-identities, also observed in humans: immunological and neuronal. (Baluška, Miller, and Reber 2022) Arguments presented by Miller, Torday, and Baluška are suggesting that biological evolution serves as a defense mechanism for the 'self'. Their findings indicate that "since all living entities display self-referential cognition, life essentially pertains to maintaining cellular homeostasis in the perpetual defense of 'self'." (Miller, Torday, and Baluška 2019) The recent EVO-EGO project, led by Watson and Buckley, explored the connectionist approach to evolutionary transitions in individuality. (Watson and Buckley 2022).

The abundance of emerging research results, especially concerning self-organization, environmental interactions, and cellular evolution, prompted Laland, Uller, Feldman, et al. to question, "Does evolutionary theory need a rethink?" (Laland et al. 2014). Their evidence leans towards a positive response, suggesting potential advancements and extensions of Modern Evolutionary Synthesis, especially at the micro-level.

The Modern Evolutionary Synthesis in its turn was a result of the extension/ fusion (merger) of Darwinian evolution with Mendelian genetics. It is sometimes referred to as the Neo-Darwinian theory.

Beyond recent suggestions to update the existing evolutionary theory of Modern Synthesis based on new data about microorganisms' behavior, Jablonka and Lamb had already earlier proposed a four-dimensional view of evolution: genetic, epigenetic, behavioral, and symbolic, anchored in cognitive learning models (Jablonka and Lamb 2014), thus extending the Modern Synthesis view (Darwin + Mendelian genetics) by epigenetic, behavioral and symbolic aspects, all of which contribute to better understanding of the processes of complexification of life through generation of new living forms. Those are not random, but goal-directed behaviors of living organisms as cognitive (information-processing) agents. Ginsburg and Jablonka's recent work explores specifically the evolution of the "sensitive soul" (Ginsburg and Jablonka, 2019), drawing inspiration from Aristotle. According to Aristotle, plants possess a nutritive soul, animals have a sensitive soul, while humans have a rational soul. Aristotle's description translates in modern terms to Maturana and Varela's concept of cognition as autopoiesis or the "realization of the living" (Maturana and Varela, 1980), where each organism, from the single living cell, possesses a degree of cognition. Ginsburg and Jablonka highlight the significance of interactions and communications between an organism and its environment, emphasizing the integral role of the genetic code's interplay with the environment in evolutionary processes.

Evolution can be interpreted as a learning process, where living entities (cells and their aggregates) adapt through environmental interactions. Aaron Sloman represents this evolutionary learning process as a meta-morphogenesis (Sloman 2013b). Watson and Szathmary (2016) ask, "How can evolution learn?" Their response is: "Connectionist models of memory and learning illustrate how basic stepwise mechanisms, tweaking relations between simple components, can craft complex system behaviors, enhancing adaptability. We introduce 'evolutionary connectionism' to acknowledge the processes through which natural selection shapes evolutionary relationships, crafting intricate system behaviors and continually refining its adaptive prowess." (Watson et al. 2016)

This learning paradigm can be expressed computationally. Analyzing the interactions between living organisms and their surroundings, Dodig-Crnkovic (2020b) posits that morphological/natural/physical computation acts as a foundational mechanism underpinning the evolutionary learning process, presenting the constraints to the generation of new forms. Levin develops a detailed picture of this agency-based process in his article "Darwin's agential materials: evolutionary implications of multiscale competency in developmental biology" (Levin 2023).

In short, after Modern Evolutionary Synthesis, Extended Evolutionary Synthesis (EES) brings new explanation mechanisms to the repertoire of evolution – biological agency on the level of cells, tissues, organs, organisms, and eco-systems, expressed epigenetic, behavioral, and symbolic aspects of evolution.

EES extends all three assumptions of Modern Synthesis:

- new variation arises through random genetic mutation [living organisms from cells up act goal-directed]
- inheritance occurs through DNA [and epigenetic factors]
- natural selection of genes is the sole cause of adaptation [additional mechanisms like learning processes]

## Connecting Anthropogenic with Biogenic and Abiotic Cognition (Humans, Animals and Machines)

Computing nature presents a common framework for understanding Anthropogenic, Biogenic, and Abiotic Cognition.

As in all of biology, nothing makes sense except in the light of evolution, (Dobzhansky 1973), and cognition as a process can only be understood in the light of evolution. Regarding abiotic systems, we compare their "cognitive behavior" with living organisms and draw conclusions.

The work of (Kaspar et al. 2021) and (Bongard and Levin 2023) suggests the direction from "cognitive/intelligent matter" towards "biological systems as evolved, overloaded, multi-scale machines". Of course, the concept of "machine" in „biological machine" is not the same as in the „mechanical machines" that we are most familiar with.

A step in-between "cognitive machine" and "biological machines" are basic elements of biological machinery like cellular signaling pathways that are plastic, (proto) cognitive systems, (Mathews et al. 2023). They provide indications that information-exchanging networks within cells also possess properties of cognitive systems such as sensitivity to input, information processing, memory, and output of result information/behavior. See references on the evolution of ribosomal protein network architectures (Timsit, Sergeant-Perthuis, and Bennequin 2021), and the study linking multimodal perception with the cellular state to decision-making in single cells (Kramer et al. 2022).

To understand the steps of the process towards increasingly competent cognitive structures, we keep in mind the role of evolution, (Dobzhansky 1973).

## Learning From Nature to Cognitively/Intelligently Compute Requires Understanding of Evolution

In the info-computational approach to cognition and intelligence, evolution is understood in the sense of extended evolutionary synthesis (Laland et al. 2015) (Ginsburg and Jablonka 2019) (Jablonka and Lamb 2014) and it is a result of interactions between natural agents, cells and their groups on a variety of levels of organization as Jablonka and Lamb argue in their "Evolution in Four Dimensions: Genetic, Epigenetic, Behavioral, and Symbolic Variation in the History of Life". These dimensions can be found on different levels of organization of life. Aaron Sloman (Sloman 2013a) defined the evolution and development of information-processing machinery in living beings as meta-morphogenesis.

## Morphological Computing

### Some Definitions Regarding Morphology

So as to support the shared view, here are some definitions.
**Morphology:** A form, shape, structure, or pattern.
**Morphogenesis**: generation of form, shape, structure, patterns, formation and transformation, patterns of formation
**Anatomy vs. Morphology:** Anatomy studies the presence of structures while morphology studies the relationships of structures. Anatomy is a subdivision of morphology, whereas morphology is a branch of biology.

## Morphological Computing in Biology

The essential property of morphological computing is that it is defined on a structure of nodes (agents) that exchange (communicate) information.

Unicellular organisms such as bacteria communicate and build swarms or films (networks) with far more advanced capabilities compared to individual organisms (nodes), through social (distributed) cognition (information exchange).

In general, groups of smaller organisms (cells) in nature cluster into bigger ones (multicellular assemblies) with differentiated control mechanisms from the sub-cellular, and cellular level to the tissue, organ, organism, and groups of organisms, and this layered organization provides information processing benefits, (Levin 2019)(Fields and Levin 2019)(Levin, Keijzer, Lyon, et al. 2021)

## Continuum of Natural Cognitive Architectures

A recent comprehensive overview of 40 years of research in cognitive architectures, (Kotseruba and Tsotsos 2020) evaluates the modeling of the core cognitive abilities in humans, but only briefly mentions biologically plausible approaches.
However, there is an important new development of biologically inspired computational models that can lead to biologically more realistic cognitive architectures.
*Unlike the vast majority of artificial cognitive architectures, that target human-level cognition, we would like to focus on the development and evolution of the continuum of natural cognitive architectures, from basal cellular up,* as studied by (Lyon et al. 2021a)(Lyon et al. 2021b) (Dodig-Crnkovic 2022c).

## Continuum of Biological Computation

The continuum of cognitive capacities from somatic cells to the nervous system and brain has been described by Levin:
*"We have previously argued that the deep evolutionary conservation of ion channel and neurotransmitter mechanisms highlights a fundamental isomorphism between developmental and behavioral processes. Consistent with this, membrane excitability has been suggested to be the ancestral basis for psychology (). Thus, it is likely that the cognitive capacities of advanced brains lie on a continuum with, and evolve from, much simpler computational processes that are widely conserved at both the functional and mechanism (molecular) levels.*

*The information processing and spatiotemporal integration needed to construct and regenerate complex bodies arise from the capabilities of single cells, which evolution exapted and scaled up as behavioral repertoires of complex nervous systems that underlie familiar examples of Selves."* (Levin 2019).

## Conclusions on  Connecting Morphological Computing and Cognition in Nature to Logic

When a cognitive agent is conceived of as a certain kind of information processor, then a pragmatic theory of cognitive agency will provide descriptions of processors of information. As mentioned before, logic is a principled account of certain aspects of practical reasoning, hence logic too is a pragmatic affair.

Cognition appears on a fundamental level in living cells in the form of basal cognition which is being extensively researched currently. We study cognition as a process of life (bio computation) on living structures (represented as bio information).

The first aspect of the connection between logic and information processing view of nature is that information processing can be viewed as a logical operation. In other words, logic can be seen as a set of rules for manipulating information, and information processing systems can be understood as physical implementations of these rules. (Benthem et al. 2001) (Benthem 2011)

On a systemic view, nature exhibits scale invariant self-organizational dynamics of energy/matter at the hierarchy of levels of organization, (Kurakin 2011).

In search for the role of logic in knowledge production, decision-making, and agency in nature, the following scheme is proposed, in a recursive (circular) manner, starting with logic:

- Logics is the intrinsic foundation of mathematics.
- Mathematics is the intrinsic logic of physics.
- Physics is the intrinsic logic of chemistry.
- Chemistry is the intrinsic logic of biology.
- Biology is the intrinsic logic of cognition.
- Cognition generates (agent's intrinsic) logic in the first place.
  An agent's intrinsic logic is a basis for creating extrinsic/externalized/shared logic. Go to 1.

The connection between logic and the information-processing view of nature is a fundamental one, with important implications for our understanding of the natural world and our ability to manipulate and control it. On this view:

- Natural processes have their own inherent logic of agency under information exchanges governed by laws of nature at different levels of organization.
- Our human-centric, language-based logic is an evolved, refined, and complex case of the logic of living organisms with basal cognition.

There is extensive empirical and theoretical work ahead to connect those phenomena with the logic of agency on different levels in nature.

The underlying mechanisms in this framework are:

- Computing universe framework – all existing nature (universe) is described as a network of networks of computational processes on informational structures.
- Nature is described as structure (information) with dynamics (computation)
- Self-organizing nature – active matter drives spontaneous processes of complexification in a universe far from thermic equilibrium.
- New computational frameworks are necessary for describing complex nature – natural computing/ unconventional computing/ interactive computing/ morphological computing.

- Cognition is seen as a result of computation of information (morphological computation, morphogenesis)
- The "mind" as a result of cognition is extended in nature (Braitenberg 2011) – embodied, embedded, and enactive.
- Info-computational formulation of The Extended Evolutionary Synthesis EES (going beyond Modern Evolutionary Synthesis by adding explanation to the processes of generation of novel biological forms and behaviors) provides generative mechanisms for the emergence of increasingly complex cognitive agents (minds) with increasingly more articulated logic. The development of the theory of evolution followed from Darwinism to Neo-Darwinism, to Modern Synthesis, and currently to the most well-developed Extended Evolutionary Synthesis which in its info-computational formulation presents the process of learning under continually changing environmental constraints.

## Acknowledgment

This paper is based on the research supported by the Swedish Research Council grant MORCOM@COGS.